\documentclass[a4paper]{jpconf}
\usepackage{graphicx}
\usepackage{amssymb}
\usepackage{amsmath}

\begin{document}

\title{Top pair production at ultra-high energies}

\author{V A Okorokov}

\address{National Research Nuclear University MEPhI (Moscow Engineering Physics Institute),
Kashirskoe highway 31, 115409 Moscow, Russia}

\ead{VAOkorokov@mephi.ru}

\begin{abstract}
The top quark, the heaviest quark and, indeed, the heaviest
elementary particle known today, constitutes a novel probe of the
long-lived medium in quark-gluon phase which, as expected, can be
produced even in light nuclei collisions at ultra-high energies.
Some distinctive features are considered for particle production
in the top sector in ultra-high energy domain. The antitop-top pair production is studied within the
quantum chromodynamics and effective field theory approach used for calculations of total partonic cross sections. Predictions for all observables are computed at NNLO in
quantum chromodynamics and at LO in effective field theory. These quantitative results can be important for both the
future collider experiments at center-of-mass energy frontier and
the improvement of the phenomenological models for development of
the cosmic ray cascades in ultra-high energy domain. Thus the
study allows the better understanding of heavy particle production
and emphasizes the exciting interrelation between the high-energy
physics on accelerators and ultra-high energy cosmic ray
measurements.
\end{abstract}

\section{Introduction}\label{sec:1}
Among the most challenging
problems for the modern physics of fundamental interactions is search for the physics beyond the Standard Model (SM) and the study of the
deconfined quark–gluon matter under extreme conditions called
also quark-gluon plasma (QGP) which can be created in subatomic
particle collisions at high enough energies.

A complete description of any physics beyond the SM requires a new fundamental theory.
From a quantum field theory (QFT) perspective potential deviations from
the SM expectations can be naturally described within the framework of the effective field theory (EFT). This approach represents a generalized parametrization of various
new physics effects on the basis of higher-dimensional operators, suppressed by a sufficiently large matching scale \cite{NPB-268-621-1986,ZPC-31-433-1986}. The top quark ($t$) sector plays an important role in searches for new physics due to largest mass of $t$ among fundamental particles of the SM and consequent its enhanced sensitivity to hypothetical new heavy particles and interactions.

The multi-TeV energies available at the Large Hadron Collider
(LHC) have already opened up the possibility to measure various high-mass
elementary particles produced in heavy ion collisions. The Higgs boson ($h$) and $t$, the heaviest elementary particles known, constitute a novel
probe of QGP. It was suggested to study the behavior of Higgs in QGP with multi-TeV collider elsewhere \cite{Okorokov-HEPFT-2014}. The study of $t$ behavior in hot environment created at ultra-high energies opens a unique possibility for investigation of a pre-equilibrium stages of space-time evolution of QGP.
In this case $h$ and $t$ can be considered as a hard probe of the properties of strongly interacting matter under extreme conditions.
The first-ever
evidence for the production of top quarks in heavy ion
interactions is recently observed in $\mbox{Pb}+\mbox{Pb}$
collisions at $\sqrt{s_{\tiny{NN}}}=5.02$ TeV
\cite{arxiv-2006.11110-2020}. The estimations obtained recently for global and geometrical parameters indicate the
creation of deconfined quark-gluon matter with large enough volume and lifetime in various type, including even light, nuclear collisions with ultra-high energy cosmic ray (UHECR) particles \cite{Okorokov-PAN-82-838-2019}.

The one of the most remarkable feature of the modern studies of UHECR is the muon deficit at energies of primary particles above on about 10 PeV \cite{EPJWoC-210-02004-2019}. In general, the production of $t$ with subsequent decay may additionally contribute in the muon yields. Furthermore the quark-gluon environment may influence on the dynamics of the processes of $t$ production and decay. That is only qualitative hypothesis which should be verified by calculations for processes with $t$ in final state in vacuum and in QGP. Thus the study of top production at ultra-high energies can be important for high-energy physics and physics of UHECR -- that is, they are of
an interdisciplinary value.

Top pair production through the strong interaction is studied within the present work. The energy range for protons in laboratory reference
system considered in the paper is
$E_{\tiny{p}}=10^{17}-10^{21}$ eV which corresponds to the c.m. collision energies $\sqrt{s_{\tiny{pp}}} \simeq 13.7-1370$ TeV, where a standard Mandelstam variable in
the form of the c.m. collision energy squared is $s_{\tiny{pp}}=2m_{\tiny{p}}(E_{\tiny{p}}+m_{\tiny{p}})$, $m_{\tiny{p}}$ is the proton mass \cite{PTEP-2020-083C01-2020}. The justification of the choice of the energy range can be found elsewhere \cite{Okorokov-PAN-82-838-2019,Okorokov-PAN-81-508-2018}. There are numerous results \cite{atlas,cms} for $\sqrt{s_{\tiny{pp}}} \simeq 14$ TeV at the right hand side value is the nominal collision energy for the LHC. Thus the present analysis mostly focuses on the ultra-high energies from $\mathcal{O}$(100 TeV) corresponded for the the Future Circular Collider (FCC)
project \cite{FCC-CDR-V3-2018} up to the $\mathcal{O}$(1 PeV) which are close to the right boundary of the energy domain for the Greisen--Zatsepin--Kuzmin limit \cite{PRL-16-748-1966,JEPTL-4-78-1966} in order of magnitude.

\section{Formalism for top pair production}\label{sec:2}
The top quark is produced in nuclear collisions predominantly in
pairs ($t\bar{t}$) through quantum chromodynamics (QCD) processes,
mostly gluon-gluon fusion at ultra-high energies $g+g \to
t+\bar{t}$. Once produced, it decays very rapidly (within an
average distance of $\sim 0.15$ fm) with almost 100\% probability
into a $W$ boson and a bottom ($b$) quark
\cite{arxiv-2006.11110-2020}. Top quark pair production is thereby
characterized by final states comprising the decay products of the
two $W$ bosons, and two $b$ jets, resulting from the hadronization
products of $b$ quarks. The dilepton final states, in which both
$W$ bosons decay into leptons ($l$) and the
corresponding neutrinos ($\nu$), are the cleanest final states for
the $t\bar{t}$ signal measurement \cite{arxiv-2006.11110-2020},
despite their relatively small branching fraction
$\mathcal{B}(t\bar{t} \to
l^{+}l^{-}\nu_{l}\bar{\nu}_{l}b\bar{b})=5.25\%$
\cite{PTEP-2020-083C01-2020}, with $l^{\pm}=e^{\pm}, \mu^{\pm}$.

Within the present work the total inclusive $t\bar{t}$ production
cross section ($\sigma_{\tiny{\mbox{tot}}}^{t\bar{t}}$) in nuclear interaction
\begin{equation}
A_{1}+A_{2} \to t+\bar{t}+X \label{eq:2.1}
\end{equation}
for the collision of the lightest nuclei $A_{1}=A_{2}=p$ involves
the following partonic subprocesses
\begin{subequations}
\begin{equation}
q+\bar{q} \to t+\bar{t}+X, \\ \label{eq:2.2.1}
\end{equation}
\begin{equation}
g+g \to t+\bar{t}+X, \\ \label{eq:2.2.2}
\end{equation}
\begin{equation}
q+g \to t+\bar{t}+X. \\ \label{eq:2.2.3}
\end{equation}\label{eq:2.2}
\end{subequations}
Using the established notations, the
$\sigma_{\tiny{\mbox{tot}}}^{t\bar{t}}$ can be written, in
particular, as follows
\begin{equation}
\sigma_{\tiny{\mbox{tot}}}^{t\bar{t}}=\sum_{ij}\int_{0}^{\beta_{\tiny{\mbox{max}}}}d\beta\,\Phi_{ij}(\beta,\mu_{F}^{2})
\hat{\sigma}_{ij}(\beta,m_{t},\mu_{F}^{2},\mu_{R}^{2}),
\label{eq:2.3}
\end{equation}

\noindent where $i,j$ run over all initial state partons studied
in the present work, $\Phi_{ij}$ is the partonic flux which is a
convolution of the densities of partons $i,j$ and includes a
Jacobian factor \cite{PRL-109-132001-2012}. The dimensionless
variable $\beta^{2}=1-\rho$, with $\rho=4m_{t}^{2}/s$, is the
squared relative velocity of the final state top quarks having pole mass
$m_{t}$ and produced at the square of the partonic center of mass
energy $s$. In general $s=x_{1}x_{2}s_{\tiny{pp}}$ with
$s_{\tiny{pp}}$ is the aforementioned square of the c.m. energy of the
colliding nuclei, namely $p$ here, $\forall\,k=1,2: x_{k}$ is the fraction of
the hadronic 4-momentum carried out by one of the incoming
partons. The fixed value $x_{1}x_{2}=1/9$ is chosen in order to
get the well-known relation between $e^{+}e^{-}$ and partonic
process $s_{e^{+}e^{-}}=s$. Here $\mu_{F,R}$ are the
renormalization and factorization scales. The total partonic
(short-distance) cross section $\hat{\sigma}_{ij}$ for the
inclusive production of a heavy quark from partons $i,j$ can be
written as an expansion in the strong coupling
\cite{NPB-303-607-1988}. For the choice of $\mu_{F}=\mu_{R}\equiv
\mu$ the NNLO partonic cross section is
\begin{equation}
\displaystyle
\hat{\sigma}_{ij}(\beta,m_{t},\mu)=\frac{\alpha_{s}^{2}}{m_{t}^{2}}\biggl\{f_{ij}^{(0)}+
\alpha_{s}\bigl[f_{ij}^{(1)}+l_{\mu}f_{ij}^{(1,1)}
\bigr]+\alpha_{s}^{2}\bigl[f_{ij}^{(2)}+l_{\mu}f_{ij}^{(2,1)}
+l_{\mu}^{2}f_{ij}^{(2,2)}\bigr]+\mathcal{O}(\alpha_{s}^{3})
\biggr\}, \label{eq:2.4}
\end{equation}
where $l_{\mu} \equiv 2\ln(\mu/m_{t})$ and $\alpha_{s}(\mu)$ is
the $\overline{\mbox{MS}}$ coupling renormalized with $N_{f}$
active flavors at scale $\mu^{2}$ \cite{PRL-109-132001-2012} and
$N_{f}=5$ in this paper. The functions $f_{ij}^{(n(,m))}$ depend
only on dimensionless parameters $\beta, \rho$. The relation
$\mu=m_{t}$ \cite{PRL-110-252004-2013} is used for the preliminary
estimations at ultra-high energies below because the NNLO partonic
cross section is available namely for this scale for gluon fusion
reaction (\ref{eq:2.2.2}) at least \cite{PRL-110-252004-2013}.
Thus the equation (\ref{eq:2.4}) can be rewritten as follows
\begin{equation}
\displaystyle
\hat{\sigma}_{ij}(\beta,m_{t})=\frac{\alpha_{s}^{2}}{m_{t}^{2}}\bigl\{f_{ij}^{(0)}+
\alpha_{s}f_{ij}^{(1)}+\alpha_{s}^{2}f_{ij}^{(2)}+\mathcal{O}(\alpha_{s}^{3})
\bigr\}. \label{eq:2.5}
\end{equation}
The functions $f_{ij}^{(0)}$ are shown elsewhere
\cite{NPB-303-607-1988} and the $f_{ij}^{(1)}$ are known exactly
through NLO \cite{NPB-824-111-2010} for the partonic subprocesses
(\ref{eq:2.2}). One can note the physically motivated fit to the
numerically integrated result from \cite{NPB-303-607-1988} are
used for the functions $f_{ij}^{(1)}$ within the present work
instead of exact complex relations \cite{NPB-824-111-2010}. As
shown in \cite{NPB-303-607-1988} the fit agrees with the
numerically integrated result to better than 1\% and this
precision seems quite reasonable and suitable for the aim of the
paper, namely for the qualitative study of $t\bar{t}$ production
in ultra-high energy domain. The NNLO functions $f_{ij}^{(2)}$ can
be found in \cite{PRL-109-132001-2012} for the partonic collisions
(\ref{eq:2.2.1}), in \cite{PRL-110-252004-2013} for
(\ref{eq:2.2.2}) and in \cite{JHEP-1301-080-2013} for $t\bar{t}$
production via quark-gluon interaction (\ref{eq:2.2.3}).

The information about the $\Phi_{ij}$ is limited and
model-dependent especially for energy range $\sqrt{s_{\tiny{pp}}} > 0.1$ PeV
due to partonic distribution functions at $\mu_{F}=m_{t}$ \cite{EPJC-79-474-2019}. That amplifies
the uncertainties for hadronic cross section (\ref{eq:2.3})
significantly. Thus the NNLO partonic cross sections
(\ref{eq:2.5}) are in the focus of the present work.

In the absence of new
resonances, effects of new physics can be described as effective
interactions of SM particles at energies below a new physics
matching scale $\Lambda$, i.e. effects beyond the SM can be
described within EFT approach. Thus the general
Lagrandian of EFT is $\mathcal{L}_{\tiny{\mbox{EFT}}}=\sum_{j=0}\mathcal{L}_{j}\Lambda^{-j}$,
where $\mathcal{L}_{\tiny{\mbox{0}}}$ is the SM Lagrangian and
$\mathcal{L}_{\tiny{\mbox{eff}}}=\sum_{j=1}\mathcal{L}_{j}\Lambda^{-j}$
-- effective part containing the effects of new physics. There is only operator at dimension five allowed by gauge invariance. Due to
the tiny neutrino masses, the scale $\Lambda$ is probably around $10^{15}$ GeV \cite{PRD-83-034006-2011}. On the other hand there are many independent dimension-six operators \cite{NPB-268-621-1986,ZPC-31-433-1986,JHEP-1010-085-2010} which provide the leading modification to SM processes at order $\Lambda^{-2}$ and simultaneously admit the existence of the significantly lower values of the scale $\Lambda$. Contributions of higher orders in the EFT expansion can be neglected if $\Lambda$ to be larger than the scale can be probed directly. Then the dominant effects are parameterized in terms of Wilson coefficients $C_{k}$ of dimension-six operators $O_{k}$ in the effective part $\mathcal{L}_{\tiny{\mbox{eff}}}=\sum_{k}\bigl(C_{k}\Lambda^{-2}{}^{+}O_{k}+\mbox{h.c.}\bigr)+\sum_{l}C_{l}\Lambda^{-2}O_{l}$, where the sum runs over all operators that involve $t$-quarks and non-hermitian operators are denoted as ${}^{+}O$ \cite{JHEP-2002-131-2020}. The lists of the dimension-six operators  for top quark production can be found elsewhere \cite{PRD-83-034006-2011,JHEP-2002-131-2020}.

As indicated above the dominant channel is (\ref{eq:2.2.2}) for top pair production through the strong interaction at ultra-high energies. Within the approach of $\mathcal{L}_{\tiny{\mbox{eff}}}$ at order $\Lambda^{-2}$ the Feynman diagrams can be found in \cite{PRD-83-034006-2011} for the gluon channel (\ref{eq:2.2.2}). The operator $O_{tG}$ changes the SM $gtt$ coupling, and also generates a new $ggtt$ interaction, $O_{G}$ affects the three-point gluon vertex in QCD and $O_{\phi G}$ generates a new diagram with an $s$-channel Higgs boson \cite{PRD-83-034006-2011}. The notation $\hat{\sigma}^{\tiny{\mbox{(0),QCD}}}_{gg}$ is used for the LO cross section of the subprocess (\ref{eq:2.2.2}) within pure QCD and $\hat{\sigma}^{\tiny{\mbox{(0),eff}}}_{gg}$ -- for the LO cross section of the partonic interaction $g+g \to t+\bar{t}$ provided by the contribution of effective part of the Lagrangian $\mathcal{L}_{\tiny{\mbox{EFT}}}$, i.e. by the higher dimension operators at order $\Lambda^{-2}$. Then the LO partonic $t\bar{t}$ production cross section due to gluon fusion (\ref{eq:2.2.2}) within EFT with leading modification to SM process ($\hat{\sigma}^{\tiny{\mbox{(0),EFT}}}_{gg}$) is \cite{PRD-83-034006-2011}
\begin{eqnarray}
\displaystyle
\hat{\sigma}^{\tiny{\mbox{(0),EFT}}}_{gg}&=&\hat{\sigma}^{\tiny{\mbox{(0),QCD}}}_{gg}+\hat{\sigma}^{\tiny{\mbox{(0),eff}}}_{gg}\nonumber\\
&=&\frac{\alpha_{s}^{2}}{m_{t}^{2}}f_{gg}^{(0)}+\mbox{Re} C_{tG}\frac{\alpha_{s}^{3/2}v}{6\Lambda^{2}}\sqrt{\frac{\pi \rho}{2s}}\biggl(8\ln\frac{1+\beta}{1-\beta}-9\beta\biggr), \label{eq:2.6}
\end{eqnarray}
where $v=246$ GeV is the vacuum expectation value (VEV) of scalar Higgs field $\phi$, the set $C_{G}=C_{\phi G}=0$ is used here because the Higgs-gluon interaction, i.e. top pair production through $g+g \to h \to t+\bar{t}$ is not considered in the present work and $O_{G}$ is strongly constrained by multi-jet production and its contribution can be neglected for $\hat{\sigma}^{\tiny{\mbox{(0),eff}}}_{gg}$ at the sensitivity reached for (\ref{eq:2.2.2}) channel \cite{JHEP-2002-131-2020}. Thus the top-gluon dipole operator $O_{tG}$ is the only contribution from the effective part of the Lagrangian $\mathcal{L}_{\tiny{\mbox{EFT}}}$ to the leading partonic subprocess (\ref{eq:2.2.2}) and one can expect a high sensitivity to $O_{tG}$ in inclusive top pair production \cite{JHEP-2002-131-2020}. The Wilson coefficient $C_{tG}$ is suggested pure real, $C_{tG} \in [0.24;0.57]$ in units of (TeV/$\Lambda)^{2}$ is used below and the range for $C_{tG}$ corresponds to the 95\% confidence level from the full global top fit with halved theoretical uncertainties \cite{JHEP-2002-131-2020}.

\begin{figure}[h!]
\begin{center}
\includegraphics[width=8.65cm]{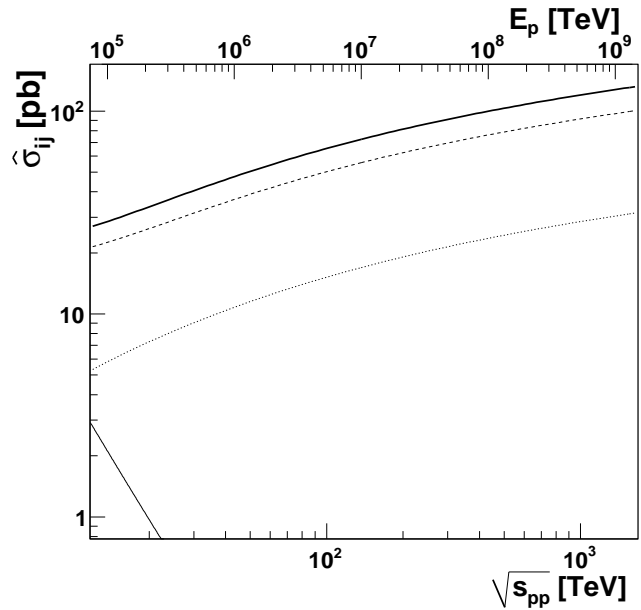}
\end{center}
\caption{\label{fig:1}Energy dependence of NNLO partonic cross sections $\hat{\sigma}_{ij}$ within QCD. The solid line shows $\hat{\sigma}_{q\bar{q}} \times 10$, the dashed line corresponds to the channel (\ref{eq:2.2.2}), the dotted line -- to the $q+g \to t+\bar{t}$ and thick solid line is the sum of the channels (\ref{eq:2.2}).}
\end{figure}

\section{Results}\label{sec:3}
Fig. \ref{fig:1} shows the energy dependence of NNLO partonic cross sections (\ref{eq:2.5}) for the channels (\ref{eq:2.2.1}) -- (\ref{eq:2.2.3}) as well as the sum of these cross sections. As expected the contribution of top pair production through $q\bar{q}$ annihilation (\ref{eq:2.2.1}) is negligible in comparison with $\hat{\sigma}_{ij}$ for the channels with incoming gluon in all energy domain under study. Futhermore $\hat{\sigma}_{q\bar{q}}$ decreases with collision energy even at NNLO level in contrast with $\hat{\sigma}_{gg}$ and $\hat{\sigma}_{qg}$ which increase smoothly with growth of $\sqrt{s_{\tiny{pp}}}$. The sum of QCD NNLO partonic cross sections for the channels (\ref{eq:2.2.1}) -- (\ref{eq:2.2.3}) reaches the value on order 0.1 nb at $\sqrt{s_{\tiny{pp}}} \gtrsim 0.5$ PeV and this value exceeds on about two times the corresponding quantity at $\sqrt{s_{\tiny{pp}}} \simeq 0.1$ PeV. These results together with the recent calculations \cite{EPJC-79-474-2019,Campbell-book-2018}
allow the qualitative expectation the total inclusive $t\bar{t}$ production cross section (\ref{eq:2.3}) at NNLO to be at level around or in order of magnitude $0.1\,\mu$b for the highest $E_{p} \simeq 10^{20}-10^{20.5}$ eV measured in UHECR. As discussed above the improving of the precision for estimations of $\Phi_{ij}$ is important for deducing the quantitative estimation of $\sigma_{\tiny{\mbox{tot}}}^{t\bar{t}}$ at ultra-high energies $\sqrt{s_{\tiny{pp}}} > 0.1$ PeV with higher accuracy and consequently for more defined physical conclusions.

\begin{figure}[h!]
\begin{center}
\includegraphics[width=8.65cm]{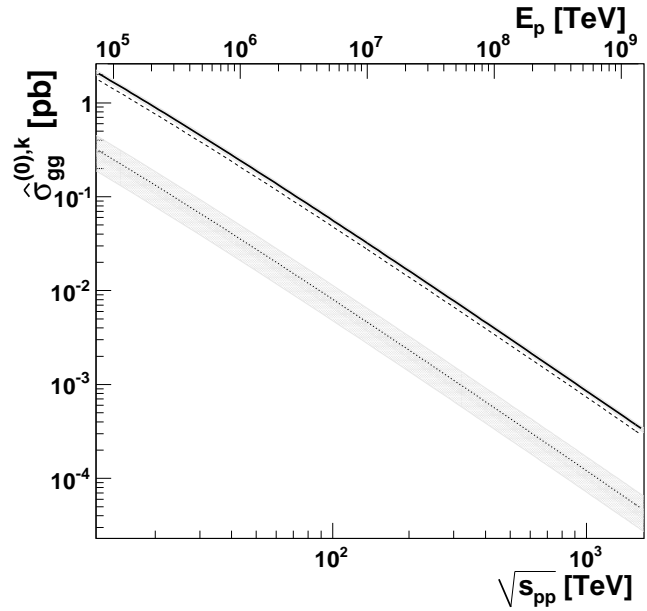}
\end{center}
\caption{\label{fig:2}Energy dependence of LO partonic cross sections $\hat{\sigma}_{ij}^{(0),k}$ within EFT for the channel (\ref{eq:2.2.2}). The dashed line corresponds to the contribution from SM ($k$=QCD), the dotted line -- to the term from the new physics effects ($k$=eff) and the solid line is the sum for the top pair production through gluon fusion in the framework of the EFT ($k$=EFT). Curves for $k$=eff and EFT are deduced for median value of $C_{tG}$, the shaded bands for these curves correspond to the spread of $\hat{\sigma}_{ij}^{(0),k}$ due to uncertainty of $C_{tG}$.}
\end{figure}

The energy dependence for channel (\ref{eq:2.2.2}) is presented in Fig. \ref{fig:2} for terms $\hat{\sigma}^{\tiny{\mbox{(0),QCD}}}_{gg}$ and $\hat{\sigma}^{\tiny{\mbox{(0),eff}}}_{gg}$ in (\ref{eq:2.6}) as well as for the sum $\hat{\sigma}^{\tiny{\mbox{(0),EFT}}}_{gg}$. As seen all LO partonic cross sections decrease with the increase of the collision energy as $\propto s^{-1}$ at qualitative level. Such behavior is well-known for the functions $f_{ij}^{(0)}$ which vanish both at threshold ($\beta \to 0$) and in high energy ($\rho \to 0$) domain \cite{Ellis-book-1996}. Fig. \ref{fig:2} shows that the contribution from the new physics effects at LO level decrease rapidly with $s$ as well. The value of $\hat{\sigma}^{\tiny{\mbox{(0),eff}}}_{gg}$ is smaller in around of order of magnitude than $\hat{\sigma}^{\tiny{\mbox{(0),QCD}}}_{gg}$ from the SM amplitudes at corresponding $s$. The LO partonic cross section from the corrections induced by new physics is $\hat{\sigma}^{\tiny{\mbox{(0),eff}}}_{gg} \sim 10$ fb at collision energy for FCC project $\sqrt{s_{\tiny{pp}}} \simeq 0.1$ PeV and $\hat{\sigma}^{\tiny{\mbox{(0),eff}}}_{gg} \sim 0.1$ fb at $\sqrt{s_{\tiny{pp}}} \simeq 1$ PeV. Despite of the relatively small values of all partonic cross sections under consideration the visible excess is observed for the aggregated cross section $\hat{\sigma}^{\tiny{\mbox{(0),EFT}}}_{gg}$ over SM term $\hat{\sigma}^{\tiny{\mbox{(0),QCD}}}_{gg}$ at confidence level (CL) larger than 1 s.d. (standard deviation) driven by the uncertainty of the Wilson coefficient $C_{tG}$. This excess is for all energy range $\sqrt{s_{\tiny{pp}}} \gtrsim 14$ TeV studied within the present work. For calculation levels higher than LO one can expect at qualitative level the increase of the difference between $\hat{\sigma}^{\tiny{(m > 0),\mbox{EFT}}}_{gg}$ and $\hat{\sigma}^{\tiny{(m > 0),\mbox{QCD}}}_{gg}$ with the growth of $s$ due to the amplification of the effects of physics beyond SM. Future theoretical developments are important for the verification of this qualitative hypothesis especially the high-order calculations for corrections induced by new physics.

The study of single top production is in the progress for ultra-high energy domain.

\section{Summary}\label{sec:4}
Top pair production through the strong interaction processes is studied for ultra-high energy domain. The partonic cross sections are evaluated to NNLO level calculation within SM. These quantities increase smoothly on collision energy within QCD for the channels with incoming gluon. The sum of QCD NNLO partonic cross sections reaches the value on order 0.1 nb at $\sqrt{s_{\tiny{pp}}} \gtrsim 0.5$ PeV. For the channel of the top pair production through gluon fusion the energy dependence is derived for LO cross sections driven by SM processes and within EFT approach with take into account dimension-six operators. LO partonic cross section decreases rapidly down to the values on order 0.1--1.0 fb at highest energies $\mathcal{O}$(1 PeV) under consideration. Despite of the relatively small values of all LO cross sections the corrections due to new physics provide visible difference between EFT and SM values for the quantities for gluon channel.

\subsection*{Acknowledgments}

The author is grateful to Prof. A. A. Petrukhin for useful
discussions. This work was supported in part within the Program for Improving
the Competitive Ability of National Research Nuclear University
MEPhI (Contract no. 02.a03.21.0005 of August 27, 2013).

\smallskip
\end{document}